# Organic and perovskite solar cells for space applications

*Ilaria Cardinaletti[1], Tim Vangerven[1], Steven Nagels[1,2], Rob Cornelissen[3], Dieter Schreurs[1], Jaroslav Hruby[1], Jelle Vodnik[3], Dries Devisscher[1], Jurgen Kesters[1], Jan D'Haen[1], Alexis Franquet[4], Valentina Spampinato[4], Thierry Conard[4], Wouter Maes[1], Wim Deferme[1,2], and Jean V. Manca[3]*

1) *Institute for Materials Research, Hasselt University, 3590 Diepenbeek, Belgium and IMEC vzw – Division IMOMEC, 3590 Diepenbeek, Belgium*
2) *Flanders Make vzw, 3920 Lommel, Belgium*
3) *X-LAB, Hasselt University, 3590 Diepenbeek, Belgium*
4) *IMEC vzw – 3000 Leuven, Belgium*

## Abstract

For almost sixty years, solar energy for space applications has relied on inorganic photovoltaics, evolving from solar cells made of single crystalline silicon to triple junctions based on germanium and III-V alloys. The class of organic-based photovoltaics, which ranges from all-organic to hybrid perovskites, has the potential of becoming a disruptive technology in space applications, thanks to the unique combination of appealing intrinsic properties (e.g. record high specific power, tunable absorption window) and processing possibilities. Here, we report on the launch of the stratospheric mission OSCAR, which demonstrated for the first time organic-based solar cell operation in extra-terrestrial conditions. This successful maiden flight for organic-based photovoltaics opens a new paradigm for solar electricity in space, from satellites to orbital and planetary space stations.

## 1. Advantages and challenges

Nearly every man-made device needs energy, most commonly in the form of electricity. This need travels along with the device, when we take it beyond the boundaries of Earth. To ensure longer lifetime and to reduce the load, solar powered satellites were introduced in the late fifties, shortly after the world wide announcement about successful solar energy harvesting[1]. PhotoVoltaics (PVs) thus allowed for truly renewable and infinitely abundant energy, the cost of which is determined only by the initial investment for the production of solar panels and, when envisioned as energy source for spacecrafts, their transport out of orbit. The cost of the latter increases quite rapidly with the mass of the object brought to space, which represents a key to the potential advantages of ultrathin solar cells. For this reason, already from the 1960s, space industry looked into the introduction of thin film $CuS_2$, CdS, and CdTe solar cells on the increasingly energy-demanding communications satellites, but eventually remained oriented on the more reliable Si[2].

Nevertheless, already in the fields of aerospace[3] and of organic and hybrid semiconductors[4,5], the specific power (W/kg) was proposed as a valid figure of merit to evaluate PV technologies for space missions. In this regard, Organic Solar Cells (OSCs) and hybrid organic-inorganic Perovskite Solar Cells (PSCs) - termed together as HOPV, Hybrid and Organic PhotoVoltaics - greatly outperform their inorganic counterparts[4,5]. They represent two novel branches of PV technologies, which saw their rise during the last decade (last few years in the case of PSCs) thanks to their potentially very low production costs. The high absorbance of the photo-active layers in HOPVs allows for efficient light collection within a few hundred nanometers of material, which leads to thicknesses one or two orders of magnitude lower than those of inorganic thin PVs. The rest of the layers making up the solar cell stacks are either as thin as or thinner than the absorbers, and the only thickness (and hence mass) limitation comes from substrate and encapsulation, which can consist of micrometers thick flexible plastic foil[4,5]. The specific power reached to date for perovskite (23 kW/kg)[4] and organic (10 kW/kg)[5] solar cells is thus over 20 or 10 times higher than what is required by some of the new missions which envision the need for lower weight and reduced deployment costs[2].

The high specific power is not the only appealing feature of these devices. The mentioned low cost fabrication originates from their intrinsic compatibility with low-temperature printing deposition techniques. They could thus be readily produced *in situ* (in/out of orbit or on a foreign planet), or transported in rolls[6]. These characteristics are quite revolutionary with respect to the PV devices currently employed by the space industry. These are folded like *origami*, to save volume, and the ensemble of hinges and structural elements makes up for most of the total mass of the final array[2]. The possibility to readily replace panels by means of printing is also of great value if we consider the heavy mechanical damage (potentially destructive) everything faces when orbiting around the Earth, where thousands of pieces of debris larger than tennis balls travel at speeds of ~10 km/s [7].

Another feature, also leading to a great potential towards high Power Conversion Efficiencies (PCEs), is the possibility to tune the energy bandgap of organic and hybrid perovskite absorbers by changing the chemical composition of the materials. Choosing for a tailored absorption window allows to optimally combine OSCs and PSCs, with each other or with inorganic PVs, in tandem devices, aiming at an increased photon collection efficiency[8].

The drawbacks holding organics and perovskites from their exploitation out of Earth are linked to the devices limited reliability, which is a paramount concern in the space industry. While on one hand the advent of new materials, processing routes, and encapsulation strategies is sure to lead towards higher stabilities, another important side of the issue lies with stability evaluation itself. Novel PV technologies are still being tested under “rooftop” degradation conditions, which do not represent the actual stress factors faced when orbiting around the Earth, for example. Space devices have to withstand unearthly harsh environments, as high energy incident radiation (mainly protons, electrons, and electromagnetic rays), a wide temperature range, vacuum, or plasma[9], depending on where they will need to operate. For example, the surface of the moon, which could represent a suitable candidate for solar energy harvesting, sees temperature variations of roughly 300 K within a few hours, and receives a flux of particles of ~$10^8$ $cm^{-2}s^{-1}$ [10]. Orbiting around the Earth together with the International Space Station would mean withstanding temperature cycles between 173 and 373 K every 45 minutes, plasmas, and a portion of the high energy charged particles radiation[7,11].

The ISOS standards[12] applied in the HOPV community are thus not sufficient to validate the degradation induced by space-related stress factors. For this reason, a few groups already started investigating the effects of high energy proton irradiation, with promising results both for all-organic[13–15] and for perovskite[16] devices. Reports on the degradation induced by wide and quickly varying temperature ranges are still missing, although insights into the effects of low operating temperatures are available from studies conducted for different aims[17]. The impact of vacuum and of plasmas on HOPVs is also unexplored, but its influence would be best countered by appropriate encapsulation and module design.

## 2. OSCAR: mission plan

Although HOPVs have a unique disruptive potential for space applications, to the best of our knowledge, these technologies have not yet been tested in real space conditions. The OSCAR[18] (Optical Sensors based on CARbon materials) mission was developed in order to demonstrate the feasibility of the use of novel generation carbon based (fully organic or hybrid organic-inorganic) solar cells for space applications. OSCAR thus fits between the huge aerospace potential of HOPVs and the lack of its testing, meaning to create a first bridge over this gap through an *in situ* study of the performance and degradation suffered by 256 solar cells (various types of OSCs and PSCs) during a stratospheric balloon flight. This pioneering investigation is, to this date, unique, because of the great challenge of reaching the stratosphere.

The experiment consisted in mounting several different HOPV devices as a load to a 35000 $m^3$ stratospheric balloon, launched in October 2016 from the Esrange Space Center, in the North of Sweden. The flight duration was limited to five hours, of which more than three in the stratosphere, reaching an altitude of 32 km (roughly 3 times higher than commercial aviation). Such an ambitious goal was attainable thanks to the support and guidance of several experts from European space-related organizations, through the REXUS/BEXUS program[19].

In order to study the performance and to screen the reliability of various materials, we selected samples of both small molecule based[20] (F4-ZnPc:$C_{60}$[21], DCV5T:$C_{60}$[22]) and polymer based[23] (PBDTTPD:$PC_{71}BM$[24], PCPDTQx(2F):$PC_{71}BM$[25]) bulk heterojunction solar cells, deposited via evaporation and spin-coating from solution, respectively. We also included a fully flexible, roll-to-roll printed, set of organic solar modules as well as spin-coated methylammonium lead triiodide perovskites ($MAPbI_3$). This wide selection of photo-active material types and deposition routes was chosen in order to cover the organic-based photovoltaics panorama as thoroughly as possible.

The flexible solar modules were purchased from InfinityPV, while the small molecule, polymer, and perovskite solar cells were prepared by the IAPP (TU/Dresden), UHasselt, and IMEC vzw, respectively. Further details on the absorbers and layer compositions are available in the Supporting Information. **Table 1** gives an overview of the performances attained by the various devices after preparation, as well as clearly indicating the total number of devices characterized during the experiment.
The selected solar cells and modules are shown in **Figure 1** as they were mounted for flight. The chosen methodology was to track the performances of the devices during flight, to obtain the evolution of the Maximum Power Point (or of other performance indicators) with time and against temperature. All data were acquired through a home built measurement unit, designed to meet the set design requirements. A detailed description of the technical aspects related to the measurement methodology and to the pre-flight tests performed can be found in the Supporting Information and in a related work [26]. The entire experiment was subject to limitations on size, weight, consumed power, and safety. It was thus mandatory to develop a dedicated and portable measurement unit which could operate in stratospheric conditions, and which would not only acquire precise current-voltage characteristics but also steadily hold all devices in place.

| Solar cell type | # | Before flight | | | | | | | | After flight | | | | | | | |
|---|---|---|---|---|---|---|---|---|---|---|---|---|---|---|---|---|---|
| | | $J_{SC}$ [mA/cm$^2$] | | $V_{OC}$ [V] | | FF [%] | | PCE [%] | | $J_{SC}$ [mA/cm$^2$] | | $V_{OC}$ [V] | | FF [%] | | PCE [%] | |
| | | Av. | St.Dev. | Av. | St.Dev. | Av. | St.Dev. | Av. | St.Dev. | Av. | St.Dev. | Av. | St.Dev. | Av. | St.Dev. | Av. | St.Dev. |
| $MAPbI_3$ | 32 | 21.4 | 0.6 | 1.0 | 0.0 | 69.7 | 3.3 | 14.6 | 1.1 | 19.3 | 4.9 | 0.9 | 0.1 | 49.4 | 20.6 | 9.3 | 5.4 |
| PBDTTPD:$PC_{71}BM$ | 32 | 10.4 | 0.9 | 0.8 | 0.1 | 57.0 | 9.3 | 4.6 | 1.4 | 9.4 | 1.1 | 0.8 | 0.1 | 49.8 | 13.0 | 3.7 | 1.3 |
| PCPDTQx(2F):$PC_{71}BM$ | 32 | 12.7 | 0.6 | 0.7 | 0.1 | 46.6 | 4.2 | 4.1 | 0.8 | 11.9 | 1.3 | 0.7 | 0.1 | 46.2 | 2.2 | 3.7 | 0.5 |
| F4-ZnPc:$C_{60}$ | 96 | 10.6 | 0.2 | 0.7 | 0.0 | 57.7 | 1.0 | 4.4 | 0.1 | 11.5 | 0.2 | 0.7 | 0.0 | 54.7 | 5.5 | 4.5 | 0.5 |
| DCV5T:$C_{60}$ | 48 | 10.7 | 0.1 | 0.9 | 0.0 | 57.1 | 1.8 | 5.4 | 0.1 | 11.7 | 0.1 | 0.9 | 0.0 | 56.7 | 0.5 | 5.9 | 0.1 |
| Flexible module | 16 | 5.0 | 0.6 | 6.0 | 0.1 | 53.8 | 0.6 | 1.6 | 0.2 | 4.1 | 0.2 | 6.0 | 0.1 | 47.6 | 2.0 | 1.2 | 0.1 |
| Total | 256 | | | | | | | | | | | | | | | | |

***Table 1***. *For each solar cell type, we list the number (#) of tested devices and their average performance parameters before and after flight, as measured under an AM1.5G simulated solar spectrum with an irradiance of 1000 W/m$^2$ (only working devices were re-measured: the number of devices included in the statistics is lower for the after flight measurements than for the before flight measurements). Since measurements were carried out in different laboratories, testing conditions might slightly vary. Due to the lack of solar simulators at the launch site and to the need for early shipment of the samples, the measurements correspond to a few months before flight and a few weeks after flight.*

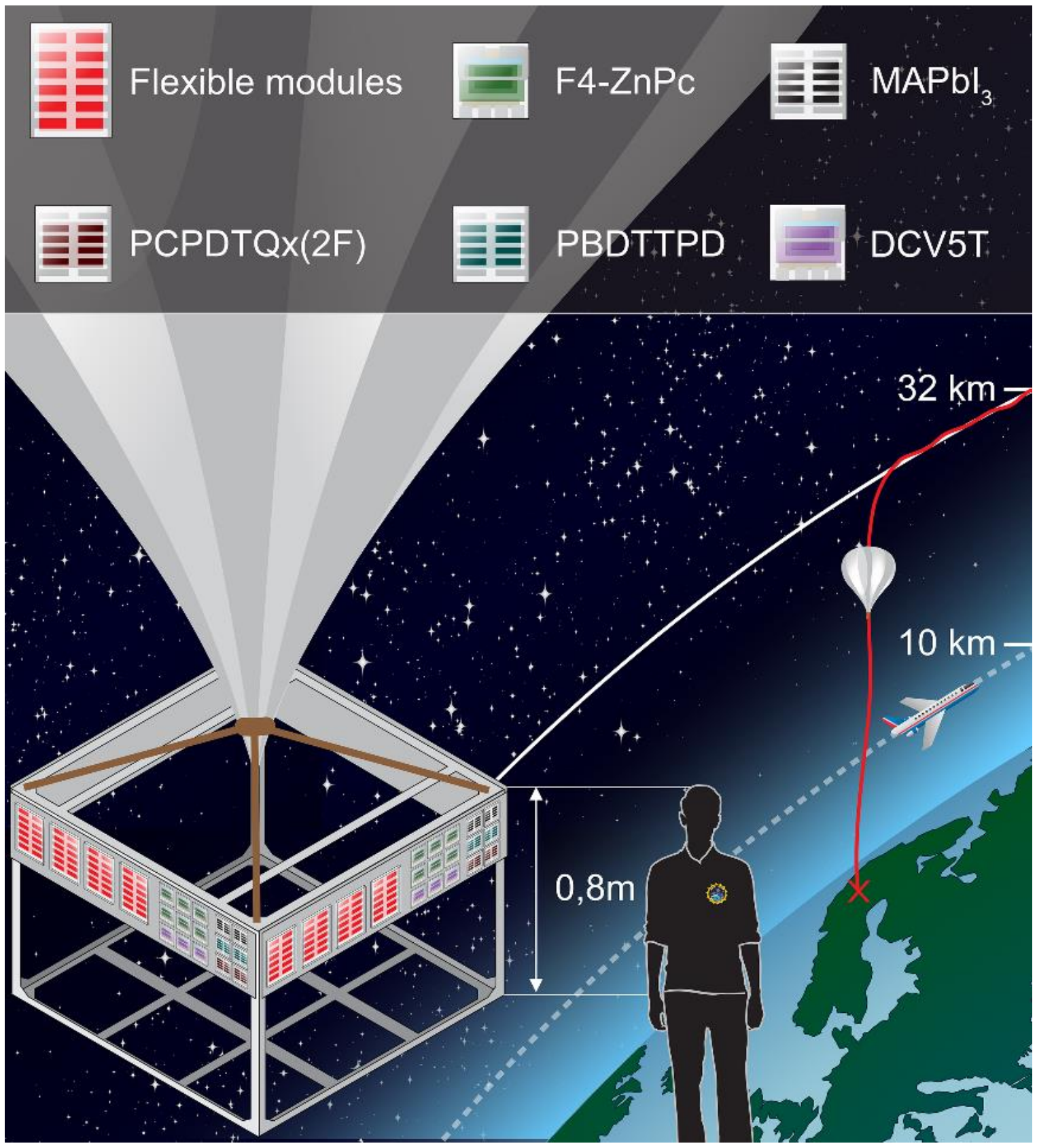

*Figure 1. Solar cells mounting structure, and schematics of the flight altitude. The experiment counted 4 panels holding solar panels, as shown. For the bulk heterojunction OSCs, devices are labeled with the donor name only.*

The atmosphere-induced spectral irradiance attenuation at 32 km of altitude can be assumed to be 1.5% of that at sea-level (details in the Supporting Information), which results in a total incident irradiance on the devices of approximately 1349 W/m$^2$. This value falls just below 1366.1 W/m$^2$, which is the current solar constant (solar irradiance at mean Earth-Sun distance from the Sun, outside Earth's atmosphere), to indicate the fact that, in terms of electromagnetic radiation, the stratosphere meaningfully represents outer space around Earth.

The radiation environment outside of our atmosphere, however, also comprises charged particles. Protons, electrons, or ions with energies as high as GeV can reach the exterior shells of the atmosphere. Most of the shielding against such charged particles happens thanks to the Earth's magnetic field, in the magnetosphere. The latter extends to several thousand km, trapping charged particles (be them originating from the sun or arriving from further in the galaxy, as galactic cosmic rays) in magnetic field lines. Because of the extension of the magnetic field of the Earth, very few protons, electrons or ions will reach the lower levels of the atmosphere. Low energy (<

50 keV) charged particles are found as far down as the ionosphere, until ~60 km of altitude. This means that, despite the less effective shielding near the North pole, the OSCAR stratospheric flight did not experience significant radiation from charged particles, if not in the form of secondary cosmic rays[11].

In terms of temperature, the tropopause (around 10 km above sea level) is the second coldest local minimum in Earth's atmosphere (~220 K). The experiment crossed this point, remained at a temperature below 230 K for roughly 30 minutes, and reached more moderate temperatures (in the range of 260-290 K) within the following hour and all throughout the phase of floatation at 32 km.

Finally, because the pressure remained higher than a few mbar, the solar cells only experienced mild vacuum during their stratospheric flight. Nevertheless, the encapsulation methods of some of the tested devices proved to be highly prone to failure, even at these moderately low pressures.

## 3. OSCAR: the results

Solar cell characteristics were acquired every 20 seconds, from a couple of hours before balloon's lift-off to roughly two and a half hours after the beginning of the float phase. Further discussion about the acquired measurements is included in the Supporting Information.
The acquisition of JV characteristics under an AM 1.5 G spectrum was not possible at the launch location, due to the lack of a solar simulator, so we could not measure the efficiency of all devices immediately before flight and immediately after flight. Nevertheless, **Table 1** shows the changes in performances between the moment of fabrication and the final re-measurement of the solar cells. The strongest drop in PCE is observed for the perovskite solar cells, mainly in terms of Fill Factor. Due to its photo-stability[27], the decay in performances of PBDTTPD:$PC_{71}BM$ is believed to be mainly linked to encapsulation defects (as described below). Speculations over the stability of the flexible modules are not possible, as the employed materials were not disclosed. However, the experienced mechanical stress (soft flexing during transportation and mounting) could have contributed to its decreased $J_{SC}$ and FF.

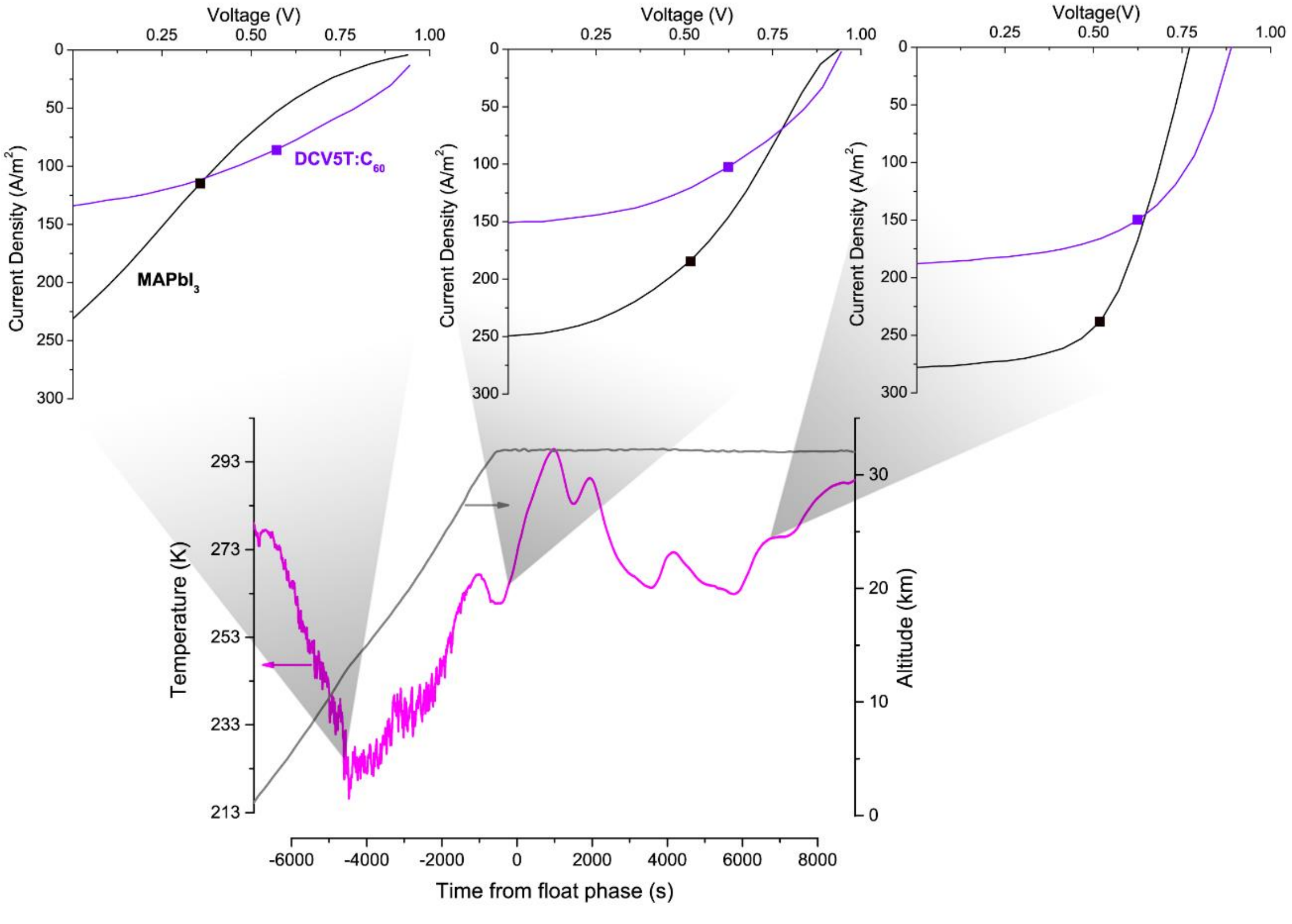

*Figure 2. Top: typical Current-Voltage curves of a perovskite ($MAPbI_3$, black) and an organic (DCV5T:$C_{60}$, purple) device at different moments. The squares on the curves represent the MPP for the shown curves. The cones point to the temperature at which the measurements were taken. The average MPP for $MAPbI_3$ was 24.84 ($\pm$9.08) W/m$^2$ at the time of the first curve, 47.21 ($\pm$19.85) W/m$^2$ at the time of the second curve, and 113.08 ($\pm$7.45) W/m$^2$ at the time of the third curve. The average MPP for DCV5T:$C_{60}$ was 47.65 ($\pm$1.87) W/m$^2$ at the time of the first curve, 63.93 ($\pm$1.62) W/m$^2$ at the time of the second curve, and 93.40 ($\pm$2.49) W/m$^2$ at the time of the third curve. Bottom: altitude (gray) and temperature (magenta) data are a courtesy of the Swedish Space Corporation and of the REXUS/BEXUS program.*

**Figure 2** shows typical in-flight Current-Voltage curves at the lowest temperature and at the beginning and end of the float phase at 32 km of altitude, for a perovskite and an organic solar cell. The high Maximum Power Point (MPP) recorded at the end of the float phase confirms that no significant degradation took place during flight. The lower Fill Factors (FFs) and short circuit currents ($J_{SC}$s) at the beginning of the flight are due to the low temperatures of operation in those moments, since charge transport in organic semiconductors is favored by mild temperatures.
A similar dependence of the performances on temperature is observed for perovskites. In this case, two possible culprits can be identified. On one hand, the charge carriers mobility within the hole selective organic small molecule will be reduced[28]. On the other hand, $MAPbI_3$ is known to undergo a smooth transition from the optimal cubic phase to the tetragonal phase at temperatures below 330 K with a consequent reduction in performance[29,30], although a decisive agreement on the magnitude of this effect is still to be reached[31]. A recent report also linked the low-temperature performance decrease to an increased interfacial recombination at the electron selecting contact[32].

All types of organic solar cells remained relatively stable during flight, as confirmed by the histograms in **Figure 3**. Here, we present the percentage of working devices for each of the tested PV technologies at three different moments: shortly after take-off, at the beginning of the float phase, and at the end of the float phase. The percentage occasionally increases with time, possibly due to a better mechanical contact, or to a momentary malfunctioning in the measurement setup.

The encapsulation of the $MAPbI_3$, the PBDTTPD:$PC_{71}BM$, and the PCPDTQx(2F):$PC_{71}BM$ proved to be less effective than that of the other solar cells. Even before launch, the devices were not “fresh”: in order to comply with the supplying schedule, all samples needed to be delivered roughly 1 month before the launch. Initial degradation had visibly taken place in devices with the mentioned active layers, due to the less effective encapsulation, which also explains why the amount of working devices in **Figure 3** was below 100% at the beginning.

The MPP of well encapsulated all-organic devices increased during the balloon’s permanence in the stratosphere, as reported in **Figure 4** for the DCV5T:$C_{60}$ solar cells. This is again due to the higher temperatures reached once the balloon started floating at 32 km of altitude. The MPP evolutions of for all the other tested materials are included in the Supplementary Information.

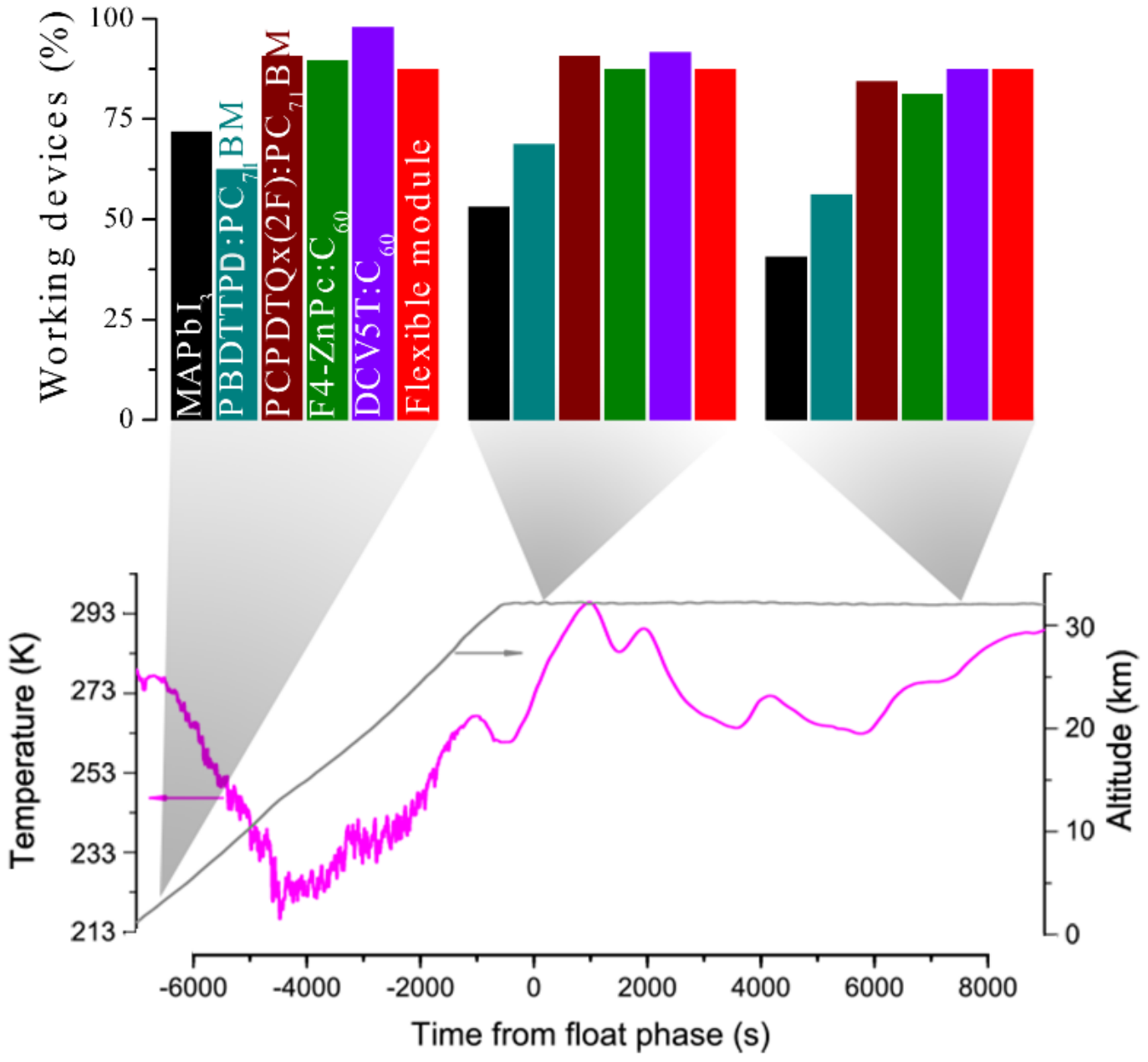

*Figure 3. Percentage of working devices throughout the flight. The histograms refer to the altitude identified by the tip of the grey shaded areas. Altitude and temperature data are a courtesy of the Swedish Space Corporation and of the REXUS/BEXUS program.*

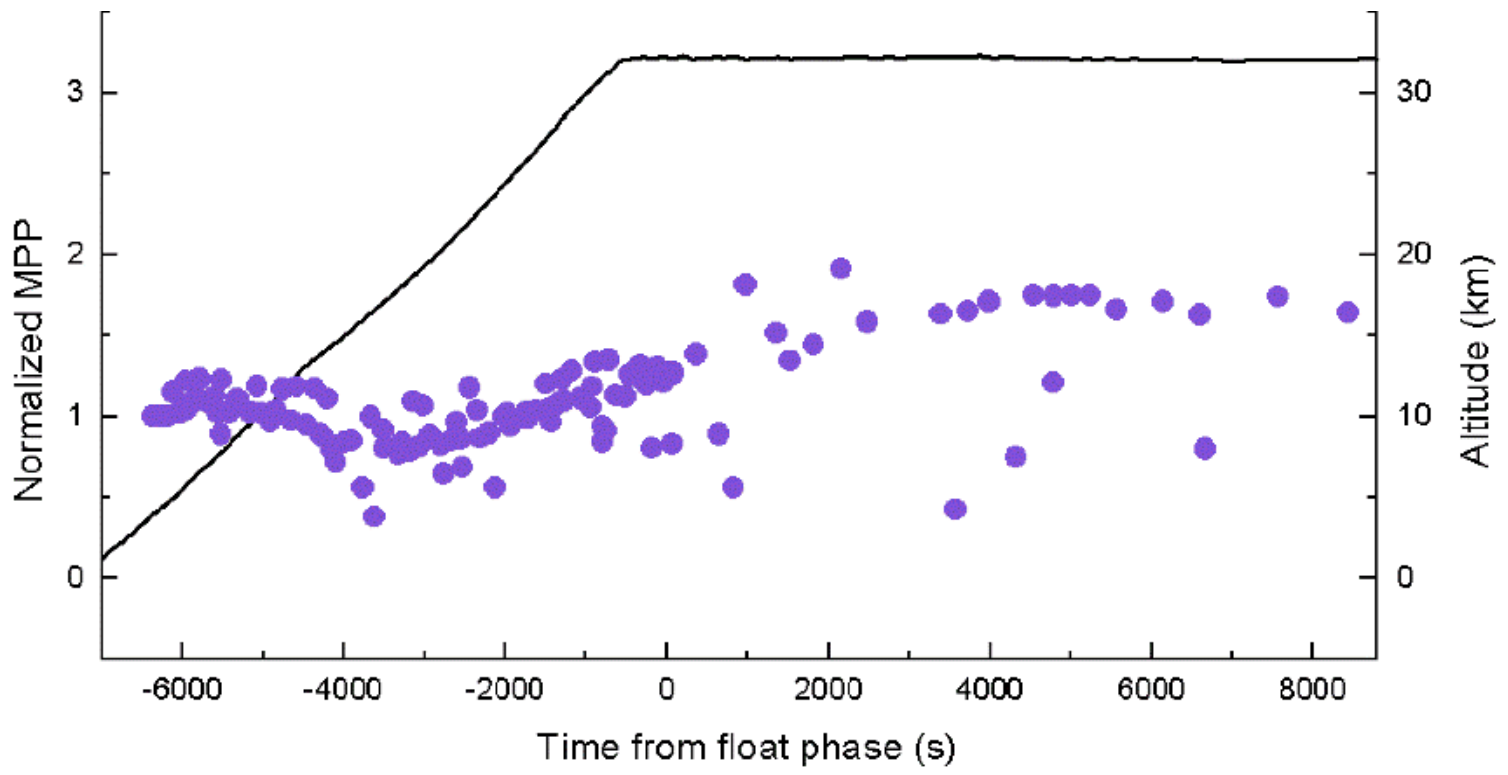

*Figure 4. Evolution of the average Maximum Power Point of DCV5T:$C_{60}$ devices during flight. The black solid line represents the altitude at each moment in time (scale on the right). Altitude data is a courtesy of the Swedish Space Center and of the REXUS/BEXUS program.*

Higher temperatures in the float phase promoted an improvement in the performances of $MAPbI_3$ solar cells as well, but the number of total working devices decreased. This was mainly due to the encapsulation, as mentioned, which completely opened up during flight in the case of two substrates. Moreover, all active layer films experienced some sort of macro-morphological degradation, visible with the naked eye as the formation of yellow bubbles in the film (**Figure 5**, top). These bubbles do not consist of perovskite phase anymore, as the bleached color clearly suggests.

The strong drop in the amount of working devices (as shown in **Figure 3**) and this evident phase change on all the perovskite layers triggered a more in-depth assessment of the cause of failure, analyzing the films where the encapsulation failed and comparing them to those where the encapsulation held until the end of the flight. Due to the limited amount of available samples, we do not possess the same (destructive) characterization for fresh devices and we refrained from breaking the encapsulation of the otherwise healthy all-organic cells and modules.

The bottom of **Figure 5** presents a schematic of the layout of the $MAPbI_3$ samples: each substrate identifies 12 solar cells, of which only 4 were measured during the flight (for technical reasons). Scanning Electron Microscopy (SEM) confirmed the formation of a large number of small protuberances distributed all over the surface of the substrates (both in correspondence with active areas and on the bare Hole Transport Layer), with a diameter of a few tens of µm. These defects are less visible over the gold electrodes, possibly due to supplementary encapsulation granted by the presence of the electrode, and on the samples, which remained encapsulated until laboratory characterization took place (see Supporting Information).

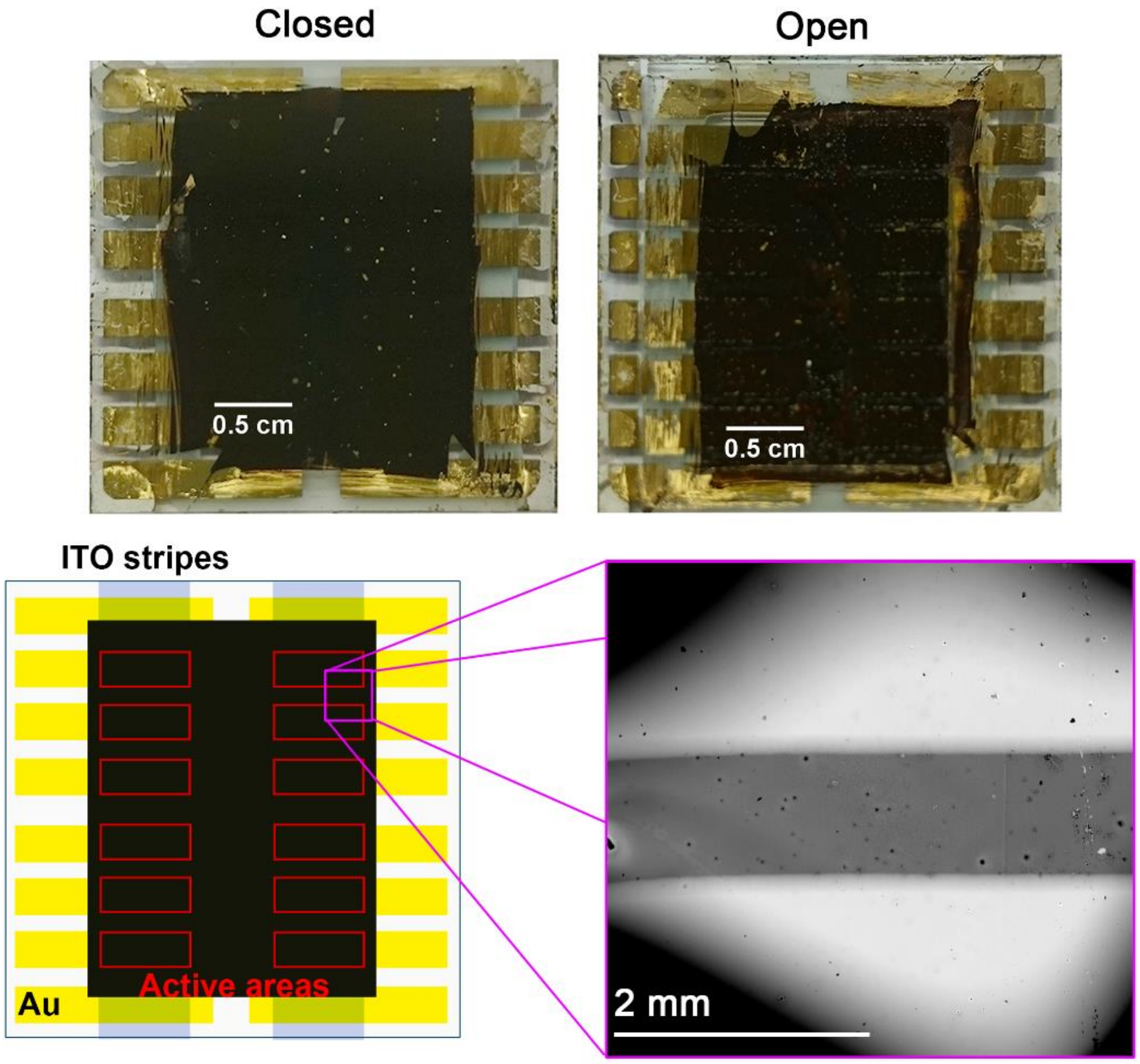

*Figure 5. Top: photographic evidence of the degradation of a perovskite device where the encapsulation remained closed (left) and failed during flight (right). Bottom: layout of the 3cm x3cm substrates, as seen through the underlying glass. The active area of the devices is 0.134 cm$^2$. SEM characterization from the selected highlighted area (magenta rectangle) shows the formation of defects (black spots) on the films.*

The films were further characterized in 3D, through depth profiling with Time Of Flight-Secondary Ion Mass Spectrometry (TOF-SIMS) operating in dual beam mode. Negative depth profiles for a few ions characteristic of the stack's components are presented in **Figure 6**, for inside and outside the mentioned protuberances. The scanned solar cell active area experienced encapsulation failure during flight.

The definition of the interfaces between the different layers within the solar cell is much clearer outside the protuberance area, where the effect of the failed encapsulation is barely visible (depth profile of "closed device" in the Supporting Information). On the other hand, the layers within the degraded areas were strongly intermixed, leading to very blurry interfaces. In particular, the Hole Selective Layer shows both upward and downward diffusion, while diffusion of the perovskite and of the Electron Selective Layer is only visible towards the substrate. The effect of the degradation is also visible in the current maps obtained with Conductive Atomic Force Microscopy (C-AFM) inside the craters created by the ion bombardment (see **Figure 6 (c)**).

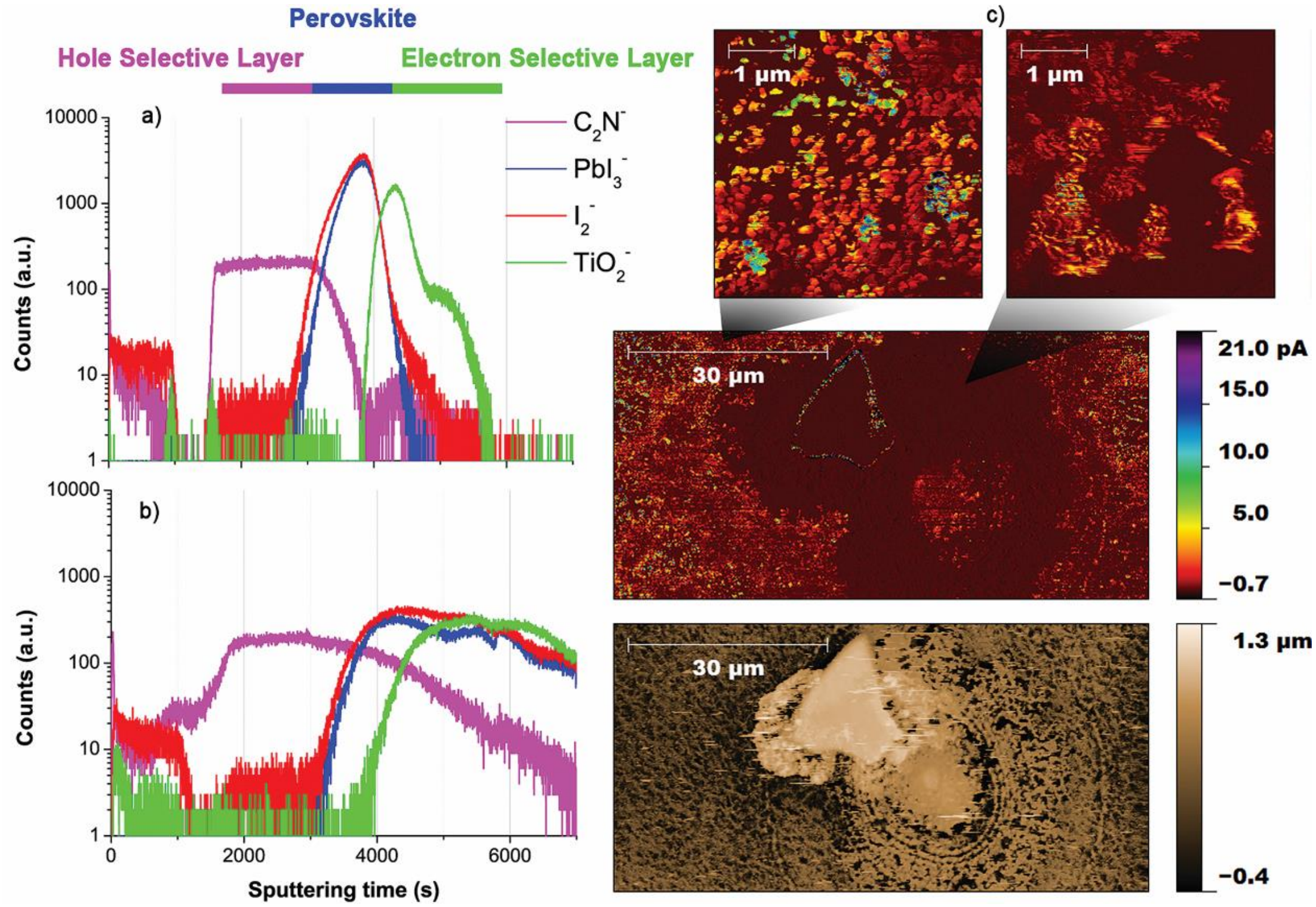

*Figure 6. TOF-SIMS depth profiles of various ions for a perovskite device of which the encapsulation failed during flight in an undamaged area (a) and within one of the protuberances visible with SEM (b). The $C_2N^-$ ions identify the organic hole selecting molecule, $PbI_3^-$ and $I_2^-$ are ascribed to the $CH_3NH_3PbI_3$ perovskite layer, and the $TiO_2^-$ represents the underlying electron selecting contact, interfacing the back electrode (See Supporting Information for details of the stack). Current maps (c) show the lower conductivity within the protuberance area. The large current scan in the middle includes the protuberance and surrounding healthy film. The corresponding height map is shown on the bottom right. On the top is the local current from smaller areas acquired near the tip of the shaded cones.*

## 4. Conclusions and outlook

Due to a unique set of intrinsic properties (*i.e.* high specific power, tunable absorption window, flexibility, foldability, …) in combination with processing possibilities in space, organic and perovskite solar cells have the potential of becoming a disruptive technology for photovoltaic energy generation in space applications. With the launch of the stratospheric mission OSCAR, organic-based solar cells where tested *in situ* in extra-terrestrial conditions for the first time.

A set of HOPV devices survived 3 hours of stratospheric flight. This exploratory outdoor degradation study in near-space environment confirms that, in principle, organic and perovskite solar cells are viable complements for space solar energy harvesting.

The present results do not offer conclusive evidence that prolonged operation in harsher conditions (as on the surface of the moon or attached to Earth orbiting satellites) will not pose significant stress on such devices. *Ex-situ* characterization of fresh devices and of devices aged at low temperatures, at high incident radiation intensity and in vacuum would provide more conclusive evidence over the causes of the observed degradation. Moreover, the mentioned scarcity of samples and the imposed delivery timeline forbade the systematic comparison of fresh, aged, and "control" samples aged in dark at room temperature, as well as the meaningful acquisition of JV curves under simulated sunlight immediately before and after the flight.

However, the very promising stability of the Maximum Power Point experienced during the flight justifies further research efforts towards indoor testing of a few of the most common space-related stress factors. A shift of paradigm could take place, for which a portion of the research effort now dedicated to the long term stabilization of solar cells against oxidizing agents might deviate towards understanding the extent of degradation introduced by extreme temperature cycling, high energy incident charged particles, mechanical "bombardment" and very low pressures.

Because of the nature of the charge transport in organic semiconductors, we envision the extreme temperatures to be the most delicate case of study, possibly leading to a large amount of materials to be unfit for very low/high temperature operation. In the case of perovskites, crystal phase transitions at precise temperatures will destroy device performances, as electrical and optical properties of the materials drastically vary upon phase change. This is true for the workhorse $MAPI_3$, as well as for other compositions (one route towards perovskite phase stabilization is to opt for mixed cations or mixed anion formulations[33,34]). Nevertheless, a careful analysis of the perovskite phase over the expected operational interval of temperature will be needed.

In conclusion, the road to walk towards this new vision is long and still vastly unexplored, which leaves ample space for future explorations.

**Acknowledgements**

The OSCAR project was developed in the framework of the REXUS/BEXUS program, supported by the Swedish National Space Board (SNSB), the Deutsches Zentrum für Luft- und Raumfahrt (DLR), the European Space Agency (ESA), and the Swedish Space Corporation.

The authors deeply thank M. De Roeve, R. Lempens, J. Soogen, K. Daniëls and J. Mertens for the technical help, and K. Wouters and M.A. Beynaerts for their support during testing. IMEC vzw is acknowledged for their availability with sample supplies and encapsulation. For the small molecule based devices thanks are due to TU/Dresden. The authors would like to thank UHasselt for additional financial support. The HERCULES Foundation financially supported the TOF-SIMS analysis. Thanks are also due to singer/song writer/astronomer Stijn Meuris for the support and for the partial use of the "Satelliet S.U.Z.Y." lyrics during pre-launch divulgation.

**Author contributions**

I.C., T.V., S.N., R.C, D.S., J.H., and J.V. were part of the team who worked on the project. I.C. and T.V. developed the solar cells testing strategy. T.V. led the team. S.N. designed and implemented the hardware and software for the measurement of solar cells. R.C. was responsible for the mechanical and thermal design. D.S. performed the pre-flight experiment validation. J.H. and J.V. designed the software, hardware, and interfaces for data transmission to the ground station. D.D. synthesized part of the photo-active materials used by J.K. to fabricate part of the solar cells (UHasselt devices). I.C. analyzed flight data and acquired the C-AFM scans. J.D'H. performed the SEM characterization. A.F., V.S., and T.C. provided the TOF-SIMS measurements. W.M., W.D. and J.V.M. supervised sections of the work. I.C. and J.V.M. wrote the manuscript, with support from all co-authors. The original idea for the study and the major endorsement came from J.V.M.

**The authors declare no competing financial interests**

**Materials and correspondence**

All correspondence and materials request should be addressed to ilaria.cardinaletti@uhasselt.be or to jean.manca@uhasselt.be .

**References**

[1] C. McLaughlin Green, M. Lomask, Vanguard - A History, 1970. doi:10.2307/1984029.

[2] S. Bailey, R. Raffaelle, Space Solar Cells and Arrays, in: Handb. Photovolt. Sci. Eng., John Wiley & Sons, Ltd, Chichester, UK, 2011: pp. 365–401. doi:10.1002/9780470974704.ch9.

[3] D. Hoffman, T. Kerslake, A. Hepp, M. Jacobs, D. Ponnusamy, Thin-film photovoltaic solar array parametric assessment, in: 35th Intersoc. Energy Convers. Eng. Conf. Exhib., American Institute of Aeronautics and Astronautics, Reston, Virigina, Virigina, 2000. doi:10.2514/6.2000-2919.

[4] M. Kaltenbrunner, G. Adam, E.D. Głowacki, M. Drack, R. Schwödiauer, L. Leonat, D.H. Apaydin, H. Groiss, M.C. Scharber, M.S. White, N.S. Sariciftci, S. Bauer, Flexible high power-per-weight perovskite solar cells with chromium oxide–metal contacts for improved stability in air, Nat. Mater. 14 (2015) 1032–1039. doi:10.1038/nmat4388.

[5] M. Kaltenbrunner, M.S. White, E.D. Głowacki, T. Sekitani, T. Someya, N.S. Sariciftci, S. Bauer, Ultrathin and lightweight organic solar cells with high flexibility, Nat. Commun. 3 (2012) 770. doi:10.1038/ncomms1772.

[6] NASA - Printable Spacecraft, (n.d.). https://www.nasa.gov/directorates/spacetech/niac/short_printable_spacecraft.html.

[7] R. Thirsk, A. Kuipers, C. Mukai, D. Williams, The space-flight environment: the International Space Station and beyond, Can. Med. Assoc. J. 180 (2009) 1216–1220. doi:10.1503/cmaj.081125.

[8] M. Filipič, P. Löper, B. Niesen, S. De Wolf, J. Krč, C. Ballif, M. Topič, CH3NH3PbI3 perovskite / silicon tandem solar cells: characterization based optical simulations, Opt. Express. 23 (2015) A263. doi:10.1364/OE.23.00A263.

[9] European Cooperation for Space Standardization, (2017).

[10] G.H. Heiken, D.T. Vaniman, B.M. French, Summary for Policymakers, in: Intergovernmental Panel on Climate Change (Ed.), Clim. Chang. 2013 - Phys. Sci. Basis, Cambridge University Press, Cambridge, 1991: pp. 1–30. doi:10.1017/CBO9781107415324.004.

[11] ECSS, ECSS-E-ST-10-04C Space environment, (2008) 198.

[12] M.O. Reese, S. a. Gevorgyan, M. Jørgensen, E. Bundgaard, S.R. Kurtz, D.S. Ginley, D.C. Olson, M.T. Lloyd, P. Morvillo, E. a. Katz, A. Elschner, O. Haillant, T.R. Currier, V. Shrotriya, M. Hermenau, M. Riede, K. R. Kirov, G. Trimmel, T. Rath, O. Inganäs, F. Zhang, M. Andersson, K. Tvingstedt, M. Lira-Cantu, D. Laird, C. McGuiness, S. (Jimmy) Gowrisanker, M. Pannone, M. Xiao, J. Hauch, R. Steim, D.M. DeLongchamp, R. Rösch, H. Hoppe, N. Espinosa, A. Urbina, G. Yaman-Uzunoglu, J.-B. Bonekamp, A.J.J.M. van Breemen, C. Girotto, E. Voroshazi, F.C. Krebs, Consensus stability testing protocols for organic photovoltaic materials and devices, Sol. Energy Mater. Sol. Cells. 95 (2011) 1253–1267. doi:10.1016/j.solmat.2011.01.036.

[13] G. Li, Y. Yang, R.A.B. Devine, Mayberr, Radiation induced damage and recovery in poly(3-hexylthiophene) based polymer solar cells, Nanotechnology. 19 (2008) 424014. doi:10.1088/0957-4484/19/42/424014.

[14] A. Kumar, R. Devine, C. Mayberry, B. Lei, G. Li, Yang, Origin of radiation-induced degradation in polymer solar cells, Adv. Funct. Mater. 20 (2010) 2729–2736. doi:10.1002/adfm.201000374.

[15] G.M. Paternò, V. Robbiano, K.J. Fraser, C. Frost, V. García Sakai, F. Cacialli, Neutron Radiation Tolerance of Two Benchmark Thiophene-Based Conjugated Polymers: the Importance of Crystallinity for Organic Avionics, Sci. Rep. 7 (2017) 41013. doi:10.1038/srep41013.

[16] F. Lang, N.H. Nickel, J. Bundesmann, S. Seidel, A. Denker, S. Albrecht, V. V. Brus, J. Rappich, B. Rech, G. Landi, H.C. Neitzert, Radiation Hardness and Self-Healing of Perovskite Solar Cells, Adv. Mater. 28 (2016) 8726–8731. doi:10.1002/adma.201603326.

[17] K. Vandewal, K. Tvingstedt, A. Gadisa, O. Inganäs, J. V Manca, Relating the open-circuit voltage to interface molecular properties of donor:acceptor bulk heterojunction solar cells, Phys. Rev. B. 81 (2010) 125204. doi:10.1103/PhysRevB.81.125204.

[18] OSCAR experiment, (n.d.). http://rexusbexus-oscar.blogspot.com.

[19] REXUS/BEXUS student experiment program, (n.d.). http://rexusbexus.net/.

[20] T. Moench, C. Koerner, C. Murawski, J. Murawski, V.C. Nikolis, K. Vandewal, K. Leo, Small Molecule Solar Cells, in: 2018: pp. 1–43. doi:10.1007/978-981-10-5924-7_1.

[21] J. Widmer, M. Tietze, K. Leo, M. Riede, Open-Circuit Voltage and Effective Gap of Organic Solar Cells, Adv. Funct. Mater. 23 (2013) 5814–5821. doi:10.1002/adfm.201301048.

[22] T. Moench, P. Friederich, F. Holzmueller, B. Rutkowski, J. Benduhn, T. Strunk, C. Koerner, K. Vandewal, A. Czyrska-Filemonowicz, W. Wenzel, K. Leo, Influence of Meso and Nanoscale Structure on the Properties of Highly Efficient Small Molecule Solar Cells, Adv. Energy Mater. 6 (2016) 1501280. doi:10.1002/aenm.201501280.

[23] L. Lu, T. Zheng, Q. Wu, A.M. Schneider, D. Zhao, L. Yu, Recent Advances in Bulk Heterojunction Polymer Solar Cells, Chem. Rev. 115 (2015) 12666–12731. doi:10.1021/acs.chemrev.5b00098.

[24] G. Pirotte, J. Kesters, P. Verstappen, S. Govaerts, J. Manca, L. Lutsen, D. Vanderzande, W. Maes, Continuous Flow Polymer Synthesis toward Reproducible Large-Scale Production for Efficient Bulk Heterojunction Organic Solar Cells, ChemSusChem. 8 (2015) 3228–3233. doi:10.1002/cssc.201500850.

[25] P. Verstappen, J. Kesters, W. Vanormelingen, G.H.L. Heintges, J. Drijkoningen, T. Vangerven, L. Marin, S. Koudjina, B. Champagne, J. Manca, L. Lutsen, D. Vanderzande, W. Maes, Fluorination as an effective tool to increase the open-circuit voltage and charge carrier mobility of organic solar cells based on poly(cyclopenta[2,1-b:3,4-b′]dithiophene-alt-quinoxaline) copolymers, J. Mater. Chem. A. 3 (2015) 2960–2970. doi:10.1039/C4TA06054G.

[26] D. Schreurs, S. Nagels, I. Cardinaletti, T. Vangerven, R. Cornelissen, J. Vodnik, J. Hruby, W. Deferme, J. Manca, Methodology of the first combined in-flight and ex-situ stability assessment of organic based solar cells for space applications, Under Rev. J. Mater. Res. (n.d.).

[27] A. Tournebize, J.-L. Gardette, C. Taviot-Guého, D. Bégué, M.A. Arnaud, C. Dagron-Lartigau, H. Medlej, R.C. Hiorns, S. Beaupré, M. Leclerc, A. Rivaton, Is there a photostable conjugated polymer for efficient solar cells?, Polym. Degrad. Stab. 112 (2015) 175–184. doi:10.1016/j.polymdegradstab.2014.12.018.

[28] I.I. Fishchuk, A.K. Kadashchuk, J. Genoe, M. Ullah, H. Sitter, T.B. Singh, N.S. Sariciftci, H. Bässler, Temperature dependence of the charge carrier mobility in disordered organic semiconductors at large carrier concentrations, Phys. Rev. B - Condens. Matter Mater. Phys. 81 (2010) 1–12. doi:10.1103/PhysRevB.81.045202.

[29] H. Zhang, X. Qiao, Y. Shen, T. Moehl, S.M. Zakeeruddin, M. Grätzel, M. Wang, Photovoltaic behaviour of lead methylammonium triiodide perovskite solar cells down to 80 K, J. Mater. Chem. A. 3 (2015) 11762–11767. doi:10.1039/C5TA02206A.

[30] L. Cojocaru, S. Uchida, Y. Sanehira, V. Gonzalez-Pedro, J. Bisquert, J. Nakazaki, T. Kubo, H. Segawa, Temperature Effects on the Photovoltaic Performance of Planar Structure Perovskite Solar Cells, Chem. Lett. 44 (2015) 1557–1559. doi:10.1246/cl.150781.

[31] T.J. Jacobsson, W. Tress, J.-P.P. Correa-Baena, T. Edvinsson, A. Hagfeldt, Room Temperature as a Goldilocks Environment for CH3NH3PbI3 Perovskite Solar Cells: The Importance of Temperature on Device Performance, J. Phys. Chem. C. 120 (2016) 11382–11393. doi:10.1021/acs.jpcc.6b02858.

[32] S. Shao, J. Liu, H. Fang, L. Qiu, G.H. ten Brink, J.C. Hummelen, L.J.A. Koster, M.A. Loi, Efficient Perovskite Solar Cells over a Broad Temperature Window: The Role of the Charge Carrier Extraction, Adv. Energy Mater. (2017). doi:10.1002/AENM.201701305.

[33] M. Saliba, T. Matsui, J.-Y. Seo, K. Domanski, J.-P. Correa-Baena, M.K. Nazeeruddin, S.M. Zakeeruddin, W. Tress, A. Abate, A. Hagfeldt, M. Grätzel, Cesium-containing triple cation perovskite solar cells: improved stability, reproducibility and high efficiency, Energy Environ. Sci. 9 (2016) 1989–1997. doi:10.1039/C5EE03874J.

[34] O.A. Syzgantseva, M. Saliba, M. Grätzel, U. Rothlisberger, Stabilization of the Perovskite Phase of Formamidinium Lead Triiodide by Methylammonium, Cs, and/or Rb Doping, J. Phys. Chem. Lett. 8 (2017) 1191–1196. doi:10.1021/acs.jpclett.6b03014.